\documentclass{article}

\usepackage{microtype}
\usepackage{graphicx}
\usepackage{subfigure}
\usepackage{booktabs} 
\usepackage{xspace}
\usepackage{amsfonts}
\usepackage{amsmath}
\usepackage[noend]{algpseudocode}
\usepackage{enumitem} 
\usepackage{nccmath}
\usepackage{amssymb,amsthm}
\algnewcommand{\LineComment}[1]{\State \(\quad\triangleright\) #1}

\renewcommand{\algorithmiccomment}[1]{\bgroup\hfill~#1\egroup}
\usepackage{algorithm}
\usepackage{listings}
\usepackage[subtle, margins=normal]{savetrees}

\usepackage{hyperref}

\setlist[enumerate]{nosep, topsep=1pt}
\setlist[itemize]{nosep, topsep=1pt}
\newcommand{\willumps}{\textsc{Willump}'s\xspace}
\newcommand{\willump}{\textsc{Willump}\xspace}

\newcommand{\toxic}{\texttt{toxic}\xspace}
\newcommand{\product}{\texttt{product}\xspace}
\newcommand{\music}{\texttt{music}\xspace}
\newcommand{\musicremote}{\texttt{music-remote}\xspace}

\newcommand{\credit}{\texttt{credit}\xspace}
\newcommand{\creditremote}{\texttt{credit-remote}\xspace}
\newcommand{\price}{\texttt{price}\xspace}

\newcommand{\purchase}{\texttt{purchase}\xspace}
\usepackage[firstpage]{draftwatermark}

\newcommand{\cost}{\text{cost}}
\newcommand{\imp}{\text{imp}}
\newcommand{\squeezeup}{\vspace{-3.0mm}}

\usepackage[accepted]{sysml2019}

\sysmltitlerunning{\willump: A Statistically-Aware End-to-end Optimizer for Machine Learning Inference}

\SetWatermarkText{
	\hspace*{4.5in}
	\raisebox{10in}{
		\includegraphics[height=0.9in]{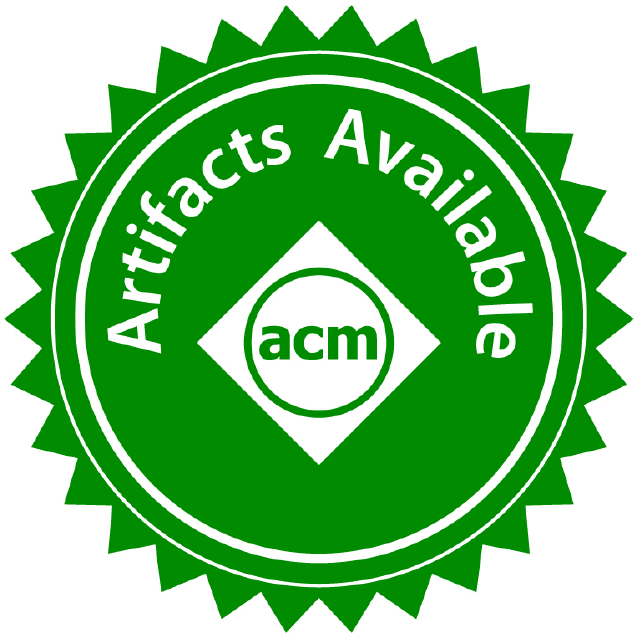}
		\includegraphics[height=0.9in]{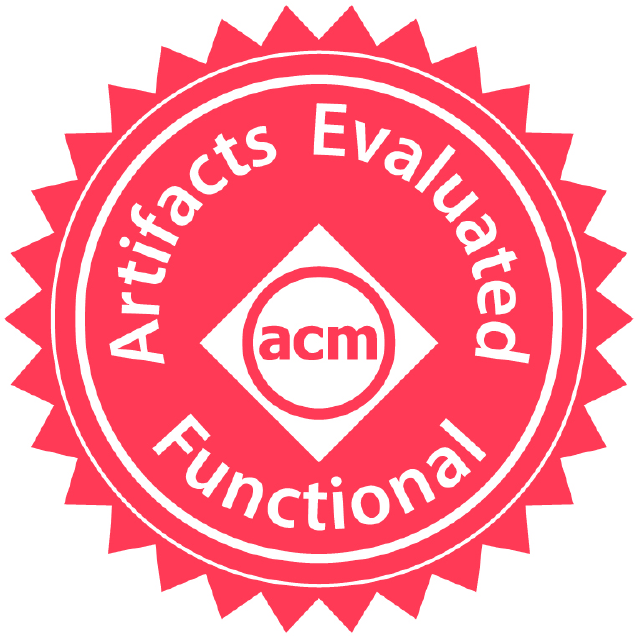}
		\includegraphics[height=0.9in]{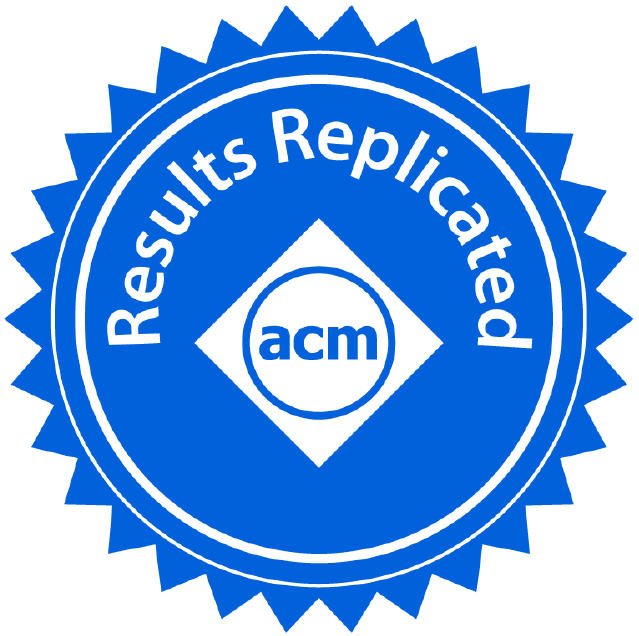}
	}
}
\SetWatermarkAngle{0}

\begin{document}
	
	\twocolumn[
	\sysmltitle{\willump: A Statistically-Aware End-to-end Optimizer for Machine Learning Inference}
	
	
	
	
	\begin{sysmlauthorlist}
		\sysmlauthor{Peter Kraft}{stan}
		\sysmlauthor{Daniel Kang}{stan}
		\sysmlauthor{Deepak Narayanan}{stan}
		\sysmlauthor{Shoumik Palkar}{stan}
		\sysmlauthor{Peter Bailis}{stan}
		\sysmlauthor{Matei Zaharia}{stan}
	\end{sysmlauthorlist}
	
	\sysmlaffiliation{stan}{Department of Computer Science, Stanford University}
	
	\sysmlcorrespondingauthor{Peter Kraft}{kraftp@cs.stanford.edu}
	
	\sysmlkeywords{Machine Learning, SysML}
	
	\vskip 0.3in
	
	\begin{abstract}
Systems for ML inference are widely deployed today, but they typically optimize ML inference workloads
using techniques designed for conventional data serving workloads and miss
critical opportunities to leverage the statistical nature of ML.
In this paper, we present \willump, an optimizer for ML inference
that introduces two statistically-motivated optimizations
targeting ML applications whose performance bottleneck is feature computation.
First, \willump automatically cascades feature computation for classification queries:
\willump classifies most data inputs using only high-value, low-cost features selected through
empirical observations of ML model performance, improving query performance by up to 5$\times$ without 
statistically significant accuracy loss.
Second, \willump accurately approximates ML top-K queries,
discarding low-scoring inputs with an automatically constructed approximate model and then
ranking the remainder with a more powerful model, improving query performance by up to 10$\times$ with minimal accuracy loss.
\willump automatically tunes these optimizations' parameters to maximize query performance while meeting an accuracy target.
Moreover, \willump complements these statistical optimizations with compiler optimizations
to automatically generate fast inference code for ML applications.
We show that \willump improves the end-to-end performance of real-world ML inference pipelines curated
from major data science competitions by up to 16$\times$ without statistically significant loss of accuracy.
\end{abstract}

	]

	
	
	\printAffiliationsAndNotice{} 
	
	\section{Introduction} \label{sec:introduction}

The importance of machine learning in modern data centers
has sparked interest in model serving systems, which perform ML inference
and serve predictions to users~\cite{crankshaw2017clipper, wang2018rafiki}.
However, these model serving systems typically approach ML inference as an extension of conventional data serving workloads,
missing critical opportunities to exploit the statistical nature of ML inference.
Most modern model serving systems, such as
Clipper \cite{crankshaw2017clipper},
Amazon Sagemaker, and Microsoft AzureML, treat ML inference as a black box and implement
generic systems optimizations such as caching and adaptive batching.
Some systems, such as Pretzel \cite{lee2018pretzel}, also apply traditional compiler optimizations
such as loop fusion.  

These optimizations are useful for ML inference applications, just as they are for web applications or database queries.
However, unlike other serving workloads, ML inference workloads have unique statistical properties that these optimizations do not leverage.
Two of these properties are:

\begin{itemize}
	\item \textbf{ML models can often be approximated efficiently for many inputs:}
    For example, the computer vision community has long used ``model cascades'' where a low-cost model
    classifies ``easy'' inputs and a higher-cost model classifies inputs where the first is uncertain,
    resulting in much faster inference with negligible change in accuracy~\cite{viola2001rapid, wang2017idk}.
	In contrast, existing multi-purpose model serving systems utilize the same logic for all data inputs.
	\item \textbf{ML models are often used for higher-level queries}, such as top-K queries.
    However, existing model serving
    systems do not optimize these query modalities.
    As we show, tailoring inference to the query (in our work, top-K queries) can improve performance.
\end{itemize}

To leverage these opportunities for optimization,
we present \willump, a system for automatically performing end-to-end optimization of ML inference workloads.
\willump targets a common class of ML inference applications:
those whose performance bottleneck is feature computation.
In these applications, a pipeline of transformations converts raw input data
into numerical features that are then used by an ML model to make predictions.
These applications are common, especially when performing ML inference over tabular data.
For example, a recent study of ML inference at Microsoft found that feature computation accounted for over
99\% of the runtime of some production ML inference applications~\cite{lee2018pretzel}.
\willump improves ML inference performance through two novel optimizations:

\paragraph{1) Automatic End-to-end Cascades:}
ML inference pipelines often compute many features for use in a model.
In classification problems, because 
ML applications are amenable to approximation, it is often possible to 
classify data inputs using only a subset of these features.
For example, in a pipeline that detects toxic online comments, we may need to compute expensive TF-IDF features
to classify some comments, but can classify others simply by checking for curse words.

However, selectively computing features is challenging because
features vary by orders of magnitude in computational cost and importance to the model,
and are often computationally dependent on one another.
Therefore, one cannot pick an arbitrary set of features 
(e.g,. the least computationally intensive) and expect to efficiently classify data inputs with them.

To address these challenges, \willump uses a cost model based on empirical observations of ML model performance
to identify important but inexpensive features.
With these features, \willump trains an approximate model that can identify and classify ``easy'' data inputs,
but \textit{cascade} ``hard'' inputs to a more powerful model.
For example, an approximate model for toxic comment classification might classify comments with
curse words as toxic but cascade other comments.
\willump automatically tunes cascade parameters to
maximize query performance while meeting an accuracy target.
The concept of cascades has a long history in the ML literature, beginning with \citet{viola2001rapid},
but to the best of our knowledge,
\willump is the first system to automatically generate feature-aware and model-agnostic cascades from input programs.
\willump's cascades  
deliver speedups of up to 5$\times$ on real-world ML inference pipelines without a statistically significant effect on accuracy.

\paragraph{2) Top-K Query Approximation:}
\willump automatically optimizes an important class of higher-level application queries:  top-K queries.
Top-K queries request a ranking of the K top-scoring
elements of an input dataset.
They are fundamentally asymmetric: predictions for high-scoring data inputs must be more precise
than predictions for low-scoring data inputs.
Existing model serving systems such as Clipper or Pretzel execute top-K queries naively, scoring
every element of the input dataset and thus wasting time generating precise predictions for low-scoring data inputs.
\willump instead leverages top-K query asymmetry by automatically constructing a computationally simple
approximate pipeline to filter out low-scoring inputs, maximizing performance while meeting a target accuracy level.
Approximation improves performance on real-world serving workloads by up to 10$\times$, with negligible impact on accuracy.

\vspace{8pt}

\willump complements end-to-end cascades and top-K query approximation with compiler optimizations.
\willump compiles a subset of Python to machine code using Weld~\cite{palkar2017weld, palkar2018evaluating},
in the process applying optimizations such as loop fusion and vectorization.
These optimizations improve query throughput by up to 4$\times$ and median query latency by up to 400$\times$.

We evaluate \willump on a broad range of pipelines curated from entries to major data science competitions
hosted by Kaggle, CIKM, and WSDM.
Overall, \willump improves query throughput by up to 16$\times$ and median query latency by up to 500$\times$.
\willumps novel optimizations contribute greatly to this performance: end-to-end cascades improve performance by up to 5$\times$ and
top-K query approximation by up to 10$\times$.
\willump also improves the performance of other model serving systems;
integrating \willump with Clipper improves end-to-end query latencies by up to 10$\times$.
All performance improvements come without statistically significant accuracy loss.

In summary, we make the following contributions:

\begin{itemize}[leftmargin=12pt]
	\item We introduce \willump, a statistically-aware end-to-end optimizer for ML inference pipelines.
	\item We describe a method for automatically cascading feature computation, improving ML inference performance by up to 5$\times$ without statistically
	significant accuracy loss.
	\item We describe a method for automatically approximating top-K queries, improving performance by up to 10$\times$ with minimal accuracy loss.
\end{itemize}

	\section{Background}\label{sec:background}

In this section, we provide background on ML inference pipelines,
cascades, and top-K queries.

\subsection{ML Inference Pipelines}

\willump optimizes ML inference applications whose performance is bottlenecked by feature computation.
In such applications, ML inference is performed by a pipeline of transformations which
receives raw input from clients, transforms it into numerical features (such as by computing statistics about a raw string input), and executes an ML model
on the features to generate predictions.
In this paper we define \textit{features} as numerical inputs to an ML model.

It is relatively common for ML inference applications to be bottlenecked by feature computation,
especially when using less expensive ML models such as linear classifiers and boosted trees.
For example, a recent study of ML inference at Microsoft found
feature computation accounted for over 99\% of the runtime of some production ML inference applications \cite{lee2018pretzel}.
Feature computation often dominates performance because it encompasses many common but relatively expensive operations in machine learning,
such as querying remote data stores~\cite{agarwal2014laser}.

Recent developments in automated machine learning (AutoML) on tabular data 
have increased the importance of feature computation.
Researchers have developed algorithms such as Google AutoML Tables~\cite{lu_2019} and 
Deep Feature Synthesis~\cite{kanter2015deep} (whose open-source implementation
is widely used~\cite{featuretools_2018}) to automatically generate ML inference pipelines
dependent on powerful but computationally expensive features.  \willump optimizes the performance of these pipelines.

\begin{figure}[t]
	\includegraphics[width=\linewidth]{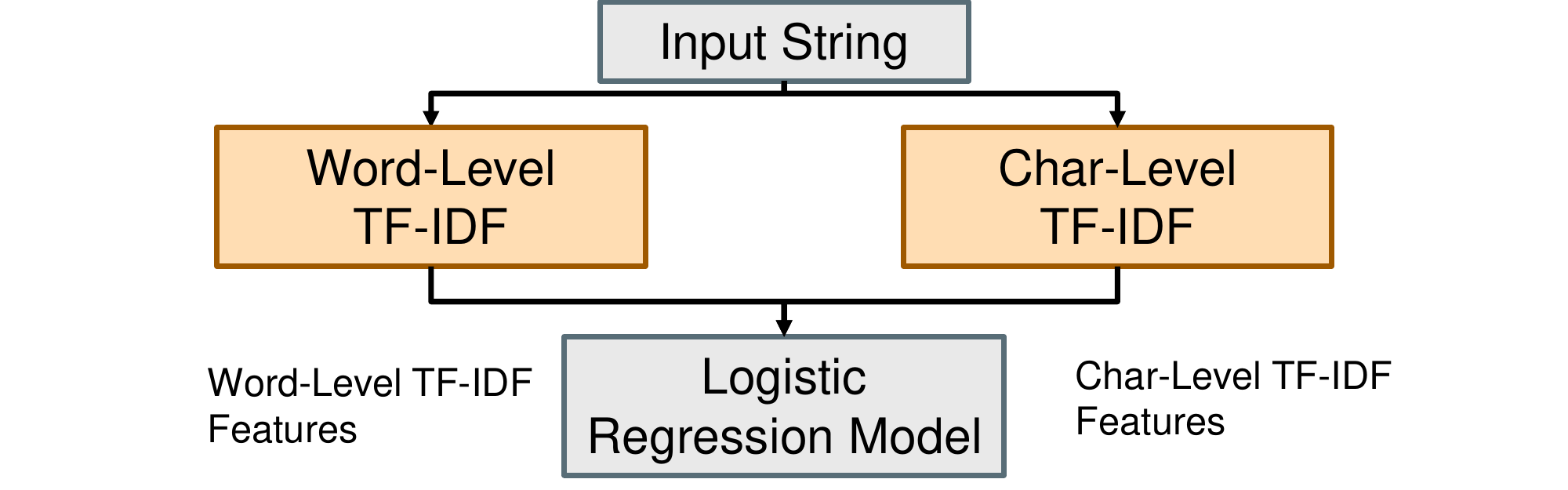}
	\caption{A simplified toxic comment classification pipeline.
		The pipeline computes word- and character-level TF-IDF features from a string
		and predicts from them with a logistic regression model.
	}
	\label{fig:sample_pipeline}
	\squeezeup
\end{figure}

We diagram an ML inference pipeline in Figure \ref{fig:sample_pipeline}.
This pipeline, which we call \toxic,
is a simplified version of one of our real-world benchmark pipelines, taken from Kaggle~\cite{tunguz}.
It predicts whether an online comment is toxic.
\toxic transforms an input string into numerical features with two TF-IDF vectorizers:  one word-level
and one character-level.  \toxic then executes a logistic regression model on these features to
predict whether the input was toxic.  In the real pipeline \toxic is based on, feature computation accounts for over 99\% of runtime.

\subsection{Cascades}

Cascades are an approximation technique for ML inference first developed for computer vision~\cite{viola2001rapid}.
ML inference is amenable to approximation because ML
models return probabilistic predictions instead of exact answers.
However, using an approximate model on every input incurs a high accuracy cost.
Cascades reduce this accuracy cost for classification problems
by leveraging differing data input difficulty.  

In most classification workloads, many data inputs are ``easy'' to classify in the sense
that a computationally simple model can accurately classify them \cite{viola2001rapid}.
Therefore, a system need not approximate every data input.
Instead, it can accurately classify easy data inputs with a computationally simple approximate model and \textit{cascade}
to a more expensive model for hard data inputs.
Existing cascades systems such as NoScope~\cite{kang2017noscope} and Focus~\cite{hsieh2018focus} 
have shown that cascades can dramatically improve ML inference performance with minimal accuracy cost,
but they are specialized to one model type.

Unlike existing cascades systems, \willump automatically optimizes an entire ML inference pipeline,
training an approximate model which depends on only a subset of the original model's features.
For example, an approximate model for \toxic might compute only word-level (and not character-level) TF-IDF features.
We call \willumps cascades optimization \textit{end-to-end} cascades.
We discuss it in more detail in Section~\ref{sec:optimizations}.

\subsection{Top-K Queries}

Top-K queries are an important class of ML inference query.
They request a ranking of the K top-scoring elements of a dataset.
Top-K queries are especially common in recommender systems~\cite{cheng2016wide}.
Database researchers have proposed several algorithms for approximating
top-K queries~\cite{theobald2004top}.  However, these algorithms require scoring functions to be
monotonic~\cite{ilyas2008survey}; this is rarely true for ML models.
Some ML recommender systems use fast retrieval models to approximate top-K queries~\cite{cheng2016wide}, but they develop these models manually.
Because ML top-K optimization is not automatic,
existing ML model serving systems such as Clipper or Pretzel do not optimize top-K queries,
instead naively scoring all elements of the input dataset.
\willump automatically approximates top-K queries, using an approximate model dependent on a subset
of the original model's features to identify and discard low-scoring inputs
before ranking remaining inputs with the original model.
We describe our approximation algorithm in detail in Section \ref{sec:optimizations}.
	\section{\willump Overview}\label{sec:design}

\begin{figure}[t]
	\includegraphics[width=\linewidth]{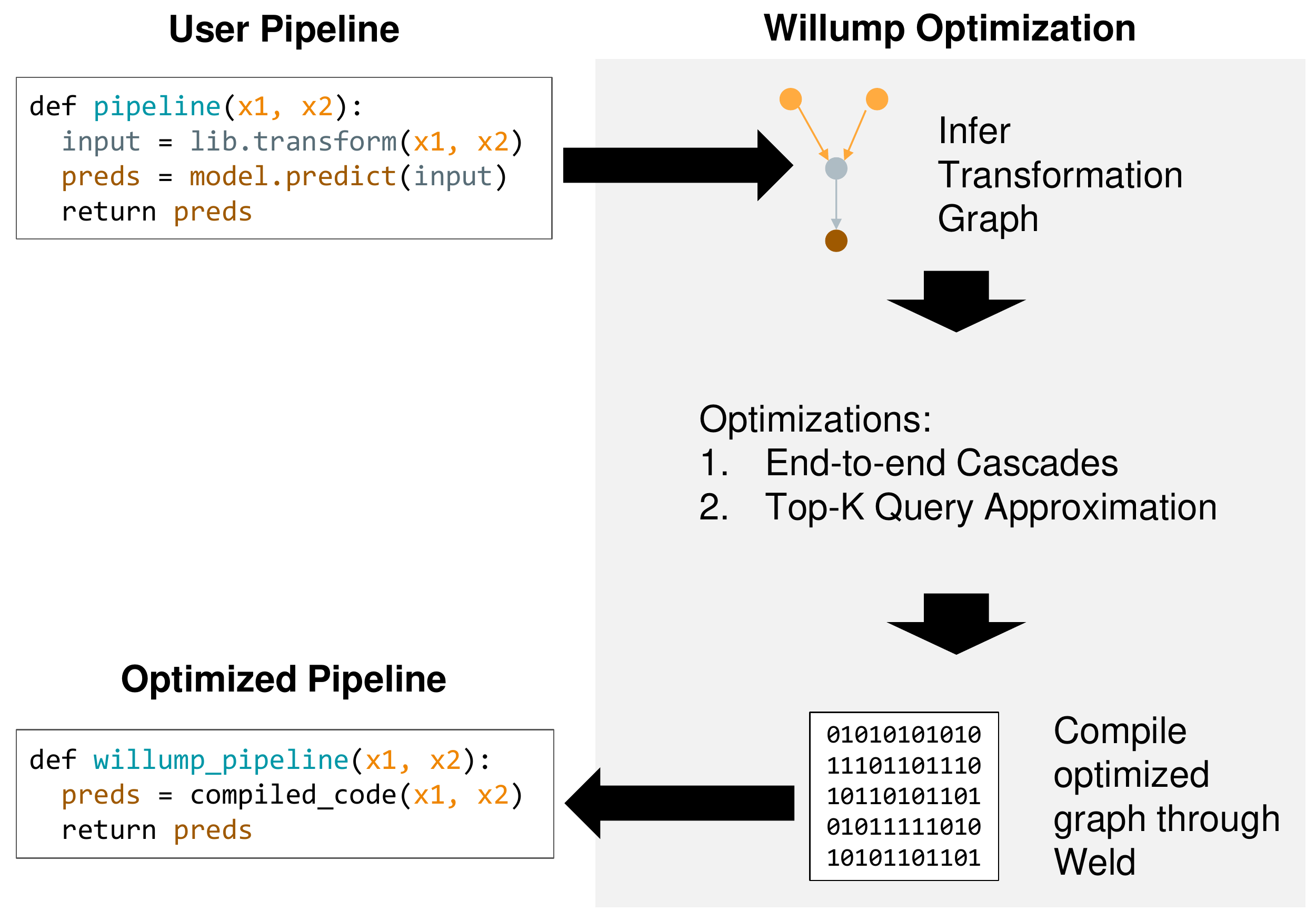}
	\caption{A diagram of \willumps architecture.  \willump infers a transformation graph from a user pipeline,
		optimizes it, compiles it through Weld, and returns an optimized pipeline.
	}
	\label{fig:architecture_diagram}
	\squeezeup
\end{figure} 

\willump is an optimizer for ML inference pipelines.
\willump users write ML inference pipelines in Python as functions from raw inputs to model predictions.
Specifically, these functions must register model training, prediction, and scoring
functions, must be written as a series of explicit Python function calls, and must represent data using
NumPy arrays, SciPy sparse matrices, or Pandas DataFrames.

\willump operates in three stages:  graph construction, optimization, and compilation.
We diagram \willumps workflow in Figure~\ref{fig:architecture_diagram}.

\paragraph{Graph Construction Stage:}  \willumps graph construction stage converts an ML inference pipeline
into a graph of transformations.
Figure \ref{fig:sample_pipeline} is an example transformation graph.
We discuss transformation graph construction in Section \ref{sec:impl-transformation-graph}.

\paragraph{Optimization Stage:}  \willumps optimization stage applies
its end-to-end cascades and top-K query approximation optimizations to the transformation graph.
We discuss these optimizations in Section \ref{sec:optimizations}.

\paragraph{Compilation Stage:}  \willumps compilation stage transforms the optimized graph back into a Python function that calls the optimized pipeline.  
In the process, it compiles some graph nodes to optimized machine code using Weld~\cite{palkar2017weld}.
We discuss graph compilation in Section \ref{sec:impl-compilation}. 

	\section{Optimizations}\label{sec:optimizations}

In this section, we describe \willumps core optimizations: 
end-to-end cascades and top-K query approximation.

\subsection{End-to-End Cascades} \label{sec:opt-cascades}

\begin{figure}[t]
	\includegraphics[width=\linewidth]{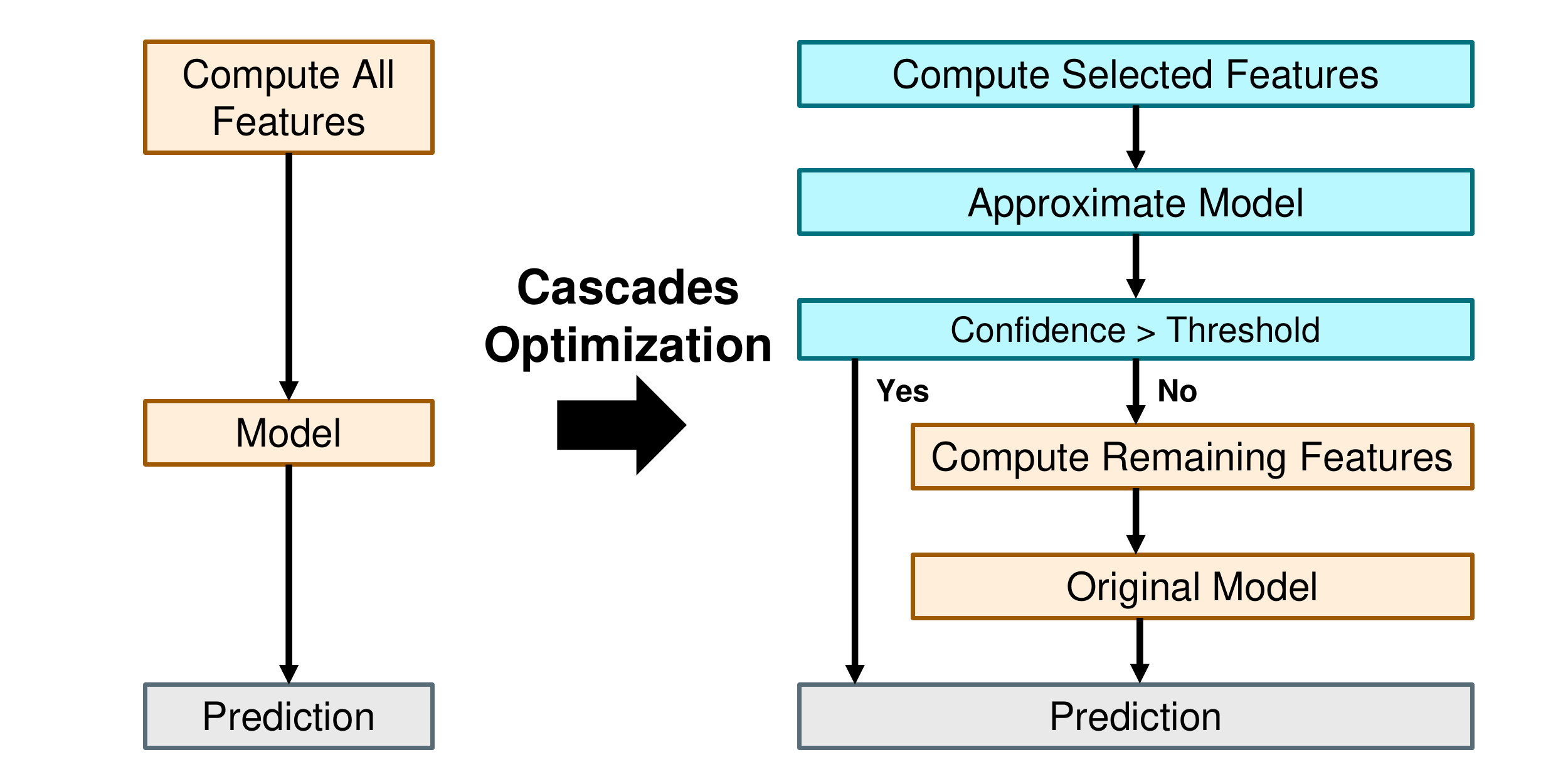}
	\caption{\willumps cascades optimization.  \willump attempts to predict
		data inputs using the approximate model, but cascades to the original model
		if the approximate model is not confident.
	}
	\label{fig:cascades_diagram}
	\squeezeup
\end{figure}

End-to-end cascades speed up ML inference pipelines that perform classification
by classifying some data inputs with an \textit{approximate model} dependent on
a subset of the original model's features. 
When using cascades, \willump first attempts to classify each data input with the approximate model.
\willump returns the approximate model's prediction if its confidence exceeds a threshold, which we
call the \textit{cascade threshold}, but otherwise
computes all remaining features and classifies with the original model.  This is shown in Figure \ref{fig:cascades_diagram}.

\willump automatically constructs end-to-end cascades from an ML inference pipeline, its training data,
and an accuracy target. First, \willump
partitions features into computationally independent groups and computes their computational cost
and importance to the model.  Then, \willump identifies several sets of computationally inexpensive but 
predictively powerful features.  For each selected set of features, \willump trains an approximate model,
chooses a cascade threshold based on the accuracy target,
and uses these to estimate the cost of accurately making predictions using cascades with those features.
\willump constructs cascades using the selected set of features that minimizes this cost.
We sketch this procedure in Algorithm~\ref{alg:cascades} and discuss it in the remainder of this section.

\begin{figure}[t]
	\includegraphics[width=\linewidth]{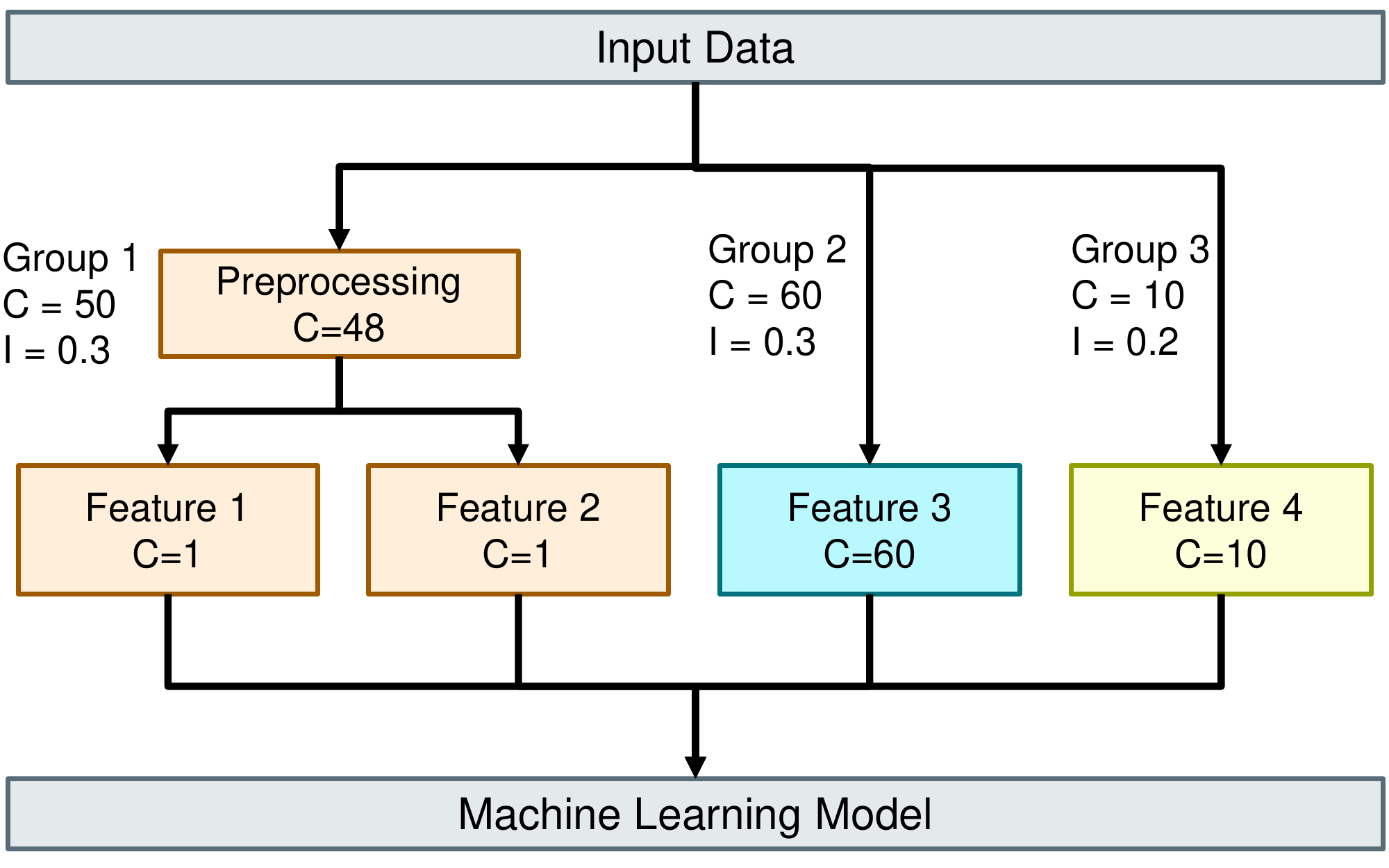}
	\caption{An example cascades optimization pipeline.  Nodes are assigned to feature groups
		and marked with their cost (C); groups are also marked with their importance (I).
		\willump might train an approximate model on Feature Groups 1 and 3 as they maximize
		sum of feature importance given maximum cost $c_{max}=60$.
	}
	\label{fig:sample_cascade_pipeline}
\end{figure}

\begin{algorithm}[t]
	\caption{Training an end-to-end cascade.}\label{alg:cascades}
	\begin{algorithmic}[1]
		\Procedure{TrainCascade}{\textbf{Graph} ${G}$, \textbf{Data} ${D}$, \textbf{Accuracy Target} $a_t$}
		\State ${T}, {H} \gets$ train\_holdout\_split$({D})$
		\State ${F} \gets \text{identify\_feature\_groups}({G})$ \Comment{\textsection\ref{sec:feature-partition}}
		\For {$f \in {F}$} \Comment{\textsection\ref{sec:feature-group-statistics}}
			\State $\imp(f) = \text{permutation\_importance}(f, {T}, {H})$
			\State $\cost(f) = \text{computational\_cost}(f, {G}, {D})$
		\EndFor
		\State $\text{min\_cost} = \infty$
		\For {$c_{max} \in {C}$}
		\LineComment{$C$ = \{ 0.1 $\cdot$ $\cost(F)$, 0.2 $\cdot$ $\cost(F)$\ldots \}}
			\State$S = \text{select\_feature\_groups}({F}, c_{max})$ \Comment{\textsection\ref{sec:query-cost-minimization}}
			\State$m= \text{train\_approximate\_model}(S, {T})$ \Comment{\textsection\ref{sec:train-approximate-models}}
			\State$t, h_S = \text{cascade\_threshold}(m, {H}, a_t)$  \Comment{\textsection\ref{sec:choose-cascade-threshold}}
			\LineComment{$t$ is the cascade threshold.}
			\LineComment{$h_S$ is the fraction of ${H}$ that $m$ can classify.}
			\State$\text{cost} = h_S\cdot\cost(S) + (1 - h_S)\cdot\cost(F)$\Comment{\textsection\ref{sec:select-optimal-features}}
			\If {$\text{cost} < \text{min\_cost}$} 
				\State best\_$S$,\, best\_$m$,\, best\_$t$ = $S,\  m,\  t$
				\State $\text{min\_cost} = \text{cost}$
			\EndIf
		\EndFor
		\State\Return{best\_$S$,\, best\_$m$,\, best\_$t$}
		\EndProcedure
	\end{algorithmic}
\end{algorithm}
\squeezeup

\subsubsection{Partitioning Features}
\label{sec:feature-partition}
The first step of constructing cascades is partitioning features into computationally independent sets, 
which we call \textit{feature groups}.
This is necessary because in most ML inference pipelines, many features are generated by the same upstream operators.
Such features are not computationally independent: it is inefficient to compute some of them
without also computing the rest. For example, in Figure \ref{fig:sample_cascade_pipeline},
Features 1 and 2 both depend on the expensive Preprocessing node, so it is 
inexpensive to compute one after computing the other.

To partition features into groups, \willump first identifies the transformation graph node that computed each feature.
It then traces those nodes' ancestors to identify their dependencies.
For example, in Figure \ref{fig:sample_cascade_pipeline}, Feature 1 depends on itself and on the Preprocessing node.
\willump assigns two features to the same feature group if the cost of their shared dependencies exceeds the cost of either feature's
unshared dependencies.
For example, in Figure \ref{fig:sample_cascade_pipeline},
it assigns Features 1 and 2 to Feature Group 1 because the cost of their shared dependency, the Preprocessing node (C$\ =48$),
exceeds the cost of either feature's unshared dependencies (C$\ =1$ for both).

\subsubsection{Computing Feature Group Statistics}
\label{sec:feature-group-statistics}
The second step of constructing cascades is to calculate two statistics for
each feature group:  its permutation importance~\cite{breiman2001random, molnar2019} and its computational cost.  \willump uses these
statistics to select features for the approximate model.
The permutation importance of a feature group is a model-agnostic measure of its value to the model's predictions~\cite{fisher2018all}.
The computational cost of a feature group is defined empirically as the amount of time it takes
to compute the feature group on a sample of the training set.

\subsubsection{Query Cost Minimization}
\label{sec:query-cost-minimization}
After identifying feature groups and computing their statistics,
\willump selects a set $S$ of computationally inexpensive but predictively
powerful feature groups from which to train an approximate model.
\willump selects the features that minimize the expected prediction time $p_t$ of an inference query
given the accuracy target $a_t$.  This time is:

\setlength{\abovedisplayskip}{1pt}
\setlength{\belowdisplayskip}{1pt}
\begin{align}
\begin{split}
\enspace p_t = \enspace h_S\cdot\cost(S) + (1 - h_S)\cdot\cost(F)
\end{split}
\label{equation:cascadeobjective}
\end{align}

Here, $F$ is the set of all features and $h_S$ is the percentage of data inputs that a cascade constructed from $S$ would classify
with the approximate model given the accuracy target $a_t$.
$h_S$ is computed using the cascade threshold, which we set later.

Unfortunately, $p_t$ is difficult to optimize directly.  At a high level, \willump approximates optimizing it by selecting several locally
optimal sets of feature groups, measuring the performance of cascades constructed from each, and choosing the best of them.

To select a locally optimal set of feature groups,
\willump considers a candidate maximum feature cost $c_{max}$.
For each candidate $c_{max}$, \willump selects the set $S$ of feature groups that minimizes $p_t$ given $\cost(S) = c_{max}$.
This set cannot score better than the set
that maximizes $h_S$ given $\cost(S) \le c_{max}$.
That set, in turn, is equivalent to the set from which
one could train an approximate model with maximum accuracy given $\cost(S) \le c_{max}$.

Unfortunately, determining approximate model accuracy for all combinations of feature groups is impractical.
Instead, \willump estimates approximate model accuracy as the sum of
the permutation importance scores of the features on which the model was trained.
Therefore, for each candidate $c_{max}$, \willump selects the set $S$
of feature groups with maximum sum of permutation importance given $\cost(S)\le c_{max}$.
This is a knapsack problem; we solve it with dynamic programming.

For each selected set of feature groups, \willump constructs cascades and measures $p_t$.
This requires training an approximate model and choosing a cascade threshold for each set.

\subsubsection{Training Approximate Models}
\label{sec:train-approximate-models}
\willump trains an approximate model from a selected set of feature groups by computing the appropriate features from the training set
and training a model of the same class as the original on them.

\subsubsection{Choosing Cascade Thresholds}
\label{sec:choose-cascade-threshold}
To choose the cascade threshold for an approximate model, \willump classifies every element in a held-out portion of the training set
using both the approximate and original models,
noting the confidences of the approximate model's predictions.  For example $i$, we call the approximate model's prediction $s_i$,
the original model's prediction $f_i$, and the approximate model's confidence $c_i$.  The cascade threshold $t$ is the lowest value
such that if we predict every data input $i$ with $s_i$ if $c_i > t$ and $f_i$ otherwise,
accuracy on the held-out set would be above the accuracy target $a_t$.

\subsubsection{Selecting Optimal Features}
\label{sec:select-optimal-features}
Using the approximate models and cascade thresholds,
\willump computes $p_t$ (from Equation \ref{equation:cascadeobjective}) for each selected set of feature groups.
\willump constructs cascades from the selected set of feature groups, and corresponding approximate model and cascade threshold, that
minimizes $p_t$.

\subsection{Top-K Query Optimization}

In top-K queries, users request a ranking of the K top-scoring elements in a dataset.
Top-K queries are fundamentally asymmetric: they must score high-scoring elements with more precision than low-scoring elements.
\willump leverages this asymmetry to approximate top-K queries by filtering out low-scoring candidates with an approximate model,
then ranking the remainder with the original model.
\willump automatically constructs these approximate models during model
training using an algorithm similar to that in Section \ref{sec:opt-cascades}.
It requires four user-specified parameters:  a distribution $\mathcal{K}$ of typical values of $K$,
a distribution $\mathcal{N}$ of typical values of the dataset size $N$,
an accuracy metric (by default precision), and
and a lower accuracy bound $a_t$.

To evaluate a top-K query, \willump scores all dataset elements with an approximate model trained on a selected set $S$
of feature groups, then ranks the $r_SK$ top-scoring
elements with the original model.
\willump chooses the set $S$ and parameter $r_S$
that minimize the expected execution time $p_t$ of a top-K query given the accuracy bound $a_t$.  This cost is:

\begin{align}
p_t = \cost(S) \cdot \overline{\mathcal{N}} + (\cost(F) - \cost(S)) \cdot r_S\overline{\mathcal{K}}
\label{equation:topk}
\end{align}

Here, $F$ is the set of all features, and $\overline{\mathcal{N}}$ and
$\overline{\mathcal{K}}$ are the means of $\mathcal{N}$ and $\mathcal{K}$.

To select a set $S$ of feature groups, \willump considers several candidate maximum feature costs $c_{max}$.
Like in Section \ref{sec:opt-cascades},
for each $c_{max}$, \willump selects the candidate set $S$ of feature groups with maximum sum of permutation 
importance given $\cost(S)\le c_{max}$.

For each candidate set $S$ of feature groups, \willump chooses a value of $r_S$, which determines the number of inputs
to be ranked by the original model.  \willump chooses the smallest value of $r_S$ that satisfies the accuracy bound $a_t$.
\willump determines this value by executing sample top-K queries on a held-out portion of the training set.
\willump first trains an approximate model and scores all elements of the held-out set.
It then draws many values $K$ and $N$ from $\mathcal{K}$ and $\mathcal{N}$.  For each pair $(K, N)$, 
it draws a sample of size $N$ from the held-out set and measures the accuracies of approximate top-K queries
run on that sample with different values of $r_S$.
\willump chooses the smallest value of $r_S$ for which accuracy is 95\% certain to be greater than $a_t$ 95\% of the time
(these thresholds can be changed by the user).  
Specifically, it chooses the smallest $r_S$
such that if we consider each sample's outcome to be a Bernoulli random variable
(where accuracy on the sample is either $\ge$ or $<$ $a_t$),
the 95\% binomial proportion confidence interval of this variable is entirely above 95\%.  

Using this procedure, \willump can compute $r_S$ and, therefore, the expected query execution time (Equation \ref{equation:topk})
for any candidate set $S$ of feature groups.
\willump selects the candidate set $S$ and corresponding $r_S$ value that minimize expected top-K query execution time given $a_t$.

	\section{\willump API and Compilation}\label{sec:implementation}

In this section we describe \willumps API and compilation procedure.

\begin{table*}[t]
	\resizebox{\textwidth}{!}{%
		\begin{tabular}{lllllrr}
			\toprule
			Benchmark & Description & Feature-Computing Operators & Prediction Type & Model  \\ \midrule
			Toxic~\cite{tunguz} & Predict whether online comment is toxic. & String processing, N-grams, TF-IDF & Classification & Linear\\
			Music~\cite{rn5l_2018} & Predict whether user will like song. & Remote data lookup, data joins & Classification &  GBDT \\
			Product~\cite{nguyen2017bagging} & Classify quality of online store listing. &String processing, N-grams, TF-IDF & Classification & Linear\\
			Instant~\cite{kikani} & Approximate a function over tabular data. & Model Stacking & Classification & Ensemble \\
			Purchase~\cite{koehrsen} & Predict customer's next purchase. & Automatically Generated Features & Classification & GBDT \\
			Price~\cite{lophuin} & Predict price of online good. & Feature encoding, string processing, TF-IDF & Regression & NN \\
			Credit~\cite{aguiar} & Predict probability a loan is defaulted. & Remote data lookup, data joins & Regression & GBDT\\
			\hline
		\end{tabular}%
	}
	\caption{Properties of \willumps benchmark workloads.}
	\label{table:datasets}
\end{table*}
\subsection{\willump API and Graph Construction} \label{sec:impl-transformation-graph}

\willump can automatically optimize an ML inference pipeline implemented as
a Python function which follows three rules.
First, the user must register model training, inference, and scoring functions that conform to a simple API;
for classification pipelines, the inference function must return a confidence metric.
This rule guarantees \willumps optimizations can be agnostic to model APIs.
Second, every statement in the pipeline must be an explicit Python function call; the final
statement, whose output is returned, must be to the model prediction function.
This rule guarantees \willump can trace data flow.
Third, the pipeline must represent all data as NumPy arrays, SciPy sparse matrices, or Pandas DataFrames.
This rule guarantees \willump can marshal data to and from Weld.  We show a simple example pipeline function using
\texttt{scikit-learn} below:

\begin{lstlisting}[language=Python]
def toxic_comment_classification(strings):
  wf = word_vectorizer.transform(strings)
  cf = char_vectorizer.transform(strings)
  return predict(model, wf, cf)
\end{lstlisting}

If an ML inference pipeline conforms to the above rules, \willump can use standard compiler techniques
to parse it into a directed acyclic
graph of transformations.  Each function called by the pipeline becomes a node in the transformation graph.
While parsing, \willump checks these functions
against a list of known functions which it can compile to Weld.
If the function is in the list, \willump marks it compilable, otherwise \willump 
leaves it in Python.

\subsection{Transformation Graph Compilation} \label{sec:impl-compilation}

After optimizing a transformation graph, \willump compiles the nodes marked compilable to optimized machine code using Weld
(leaving the remaining nodes in Python)
and packages them back into a Python function.  Weld~\cite{palkar2018evaluating}
is an intermediate representation (IR) and compiler
for data processing operations.
Weld implements many compiler optimizations over its IR, including loop fusion, data structure preallocation,
and vectorization.  These are similar to the compiler optimizations implemented by some existing model serving
systems such as Pretzel~\cite{lee2018pretzel}.
The set of operators \willump can compile to Weld is extensible; users can add Weld IR for custom operators.
\willump compiles graphs in four stages:

\squeezeup
\paragraph{Sorting:}  In the first stage of compilation, \willump sorts the transformation graph into an ordered list of nodes
which minimizes the number of transitions between compilable nodes (which are executed in Weld) and
Python nodes.  This is desirable because each transition requires
marshaling data between languages and because the Weld optimizer can apply end-to-end optimizations like loop fusion over large Weld blocks.  
\willump sorts the graph topologically, then heuristically minimizes the number of transitions by 
moving each Python node to the earliest allowable location.
We observe that this is effective in our evaluation
because Python nodes are typically either preprocessing nodes (which execute before any Weld nodes)
or parts of the model (which execute after all Weld nodes).

\squeezeup
\paragraph{Code Generation:}  In the second stage of compilation, \willump compiles every compilable node to the Weld IR.
\willump compiles nodes using parameterized Weld templates.  For example, it compiles a TF-IDF vectorization node
with a template containing a TF-IDF implementation written in Weld and 
parameterized to use several different tokenizers, n-gram ranges, and norms.  After generating Weld code,
\willump coalesces Weld and uncompiled Python code segments together, creating
\textit{blocks} of Weld and Python code.
\squeezeup
\paragraph{Drivers:}  In the third stage of compilation, \willump generates a \textit{driver},
a Python extension that calls into Weld,
for each block of Weld code.
\willump identifies each block's input and output variables,
then generates C++ driver code that
marshals each block's input variables into Weld data structures,
executes its Weld code, and marshals its output variables into Python data structures.
To minimize driver latency, we developed several new Weld types for \willump, including
dataframe and sparse matrix types, which drivers can create in $O(1)$ time from their Python equivalents.
\squeezeup
\paragraph{Compilation:}  In the final stage of compilation, \willump combines the
Python blocks and Weld drivers into a Python program.
\willump compiles drivers into Python C extensions,
inserts calls to them in the appropriate places,
packages the resulting all-Python program as a Python function, and returns it.

	\section{Evaluation} \label{sec:evaluation}

We evaluate \willump and its optimizations on seven high-performing entries to major data science competitions
at CIKM, Kaggle, and WSDM.
We demonstrate that:

\begin{enumerate}
	\item \willump improves batch inference throughput by up to 16$\times$ and point query latency by up to 500$\times$.
	Of this speedup, up to 5$\times$ comes from end-to-end cascades, and the rest from compilation.
	
	\item \willump improves top-K query performance by up to 31$\times$.
	Of this speedup, up to 10$\times$ comes from top-K query approximation, and the rest from compilation.
	
	\item \willump improves the performance of other model serving systems.
	\willump improves end-to-end Clipper performance by up to 9$\times$.
	Moreover, \willumps optimizations improve performance over compiler optimizations similar
	to Pretzel's\footnote{We cannot compare against Pretzel directly as its code and benchmarks are not publicly available}.
\end{enumerate}

\subsection{Experimental Setup}

We ran all benchmarks on an n1-standard-8 Google Cloud instance with four Intel Xeon CPUs running at 2.20 GHz with 30 GB of RAM.
We stored remote data tables on a Redis 3.2.6 server on a GCP instance with a single Intel Xeon
CPU running at 2.20 GHz with 30 GB of RAM.

\subsection{Benchmarks}

\begin{figure}[t]
	\includegraphics[width=\linewidth]{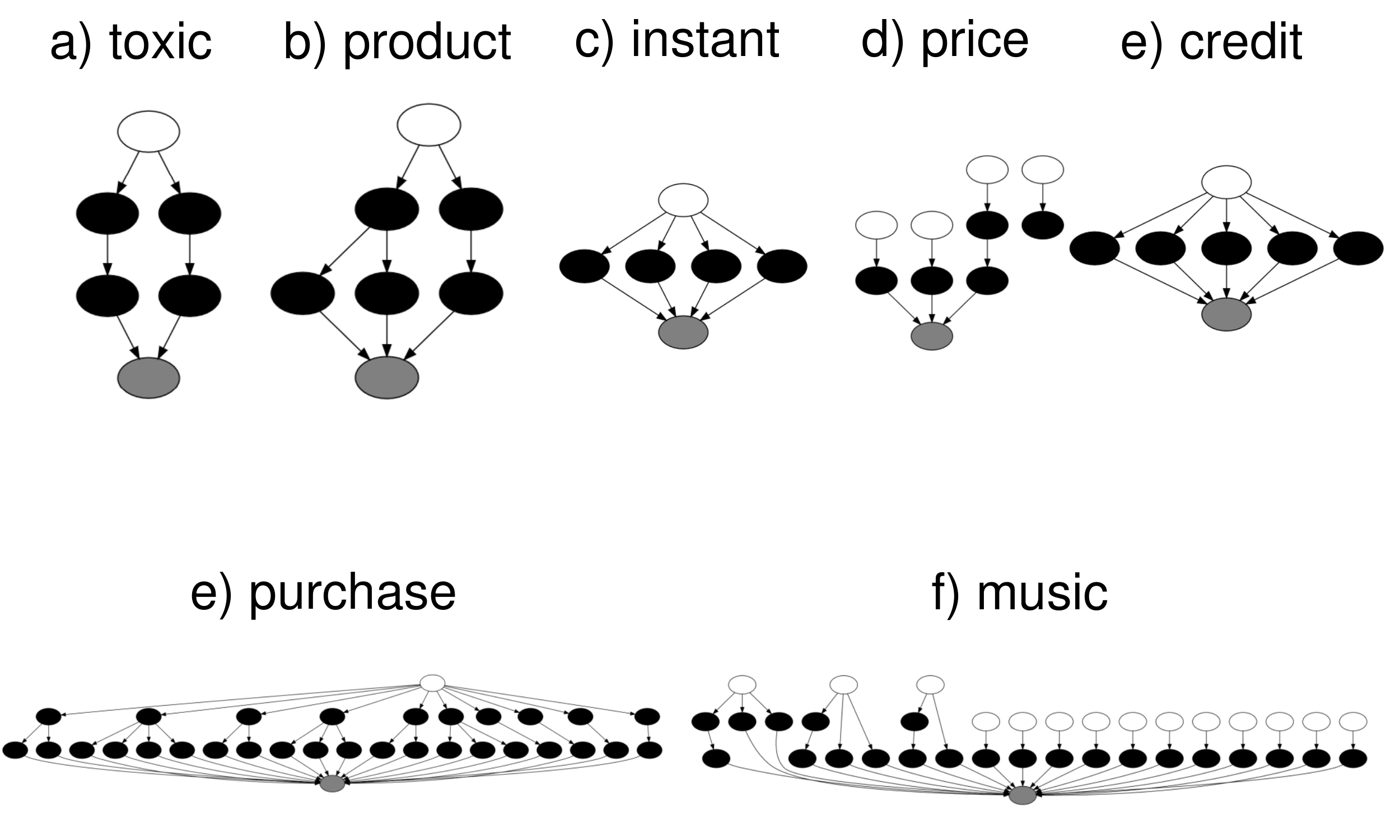}
	\caption{Transformation graphs of all benchmark workloads.
		Inputs are in white, models in gray,
		and transformation nodes in black.}
	\label{fig:transformation_graphs}
	\squeezeup
\end{figure}

We benchmark \willump on the seven real-world workloads described in Table \ref{table:datasets}, 
all curated from entries in major data science competitions.
All benchmarks are single-threaded.
One benchmark, \purchase, uses features automatically generated by the Deep Feature Synthesis algorithm~\cite{kanter2015deep};
these features consist largely of aggregations such as averages and counts over relational data.
Two benchmarks, \credit and \music, query data from stores which can be located either locally or remotely.
We query \music remotely in all benchmarks;
we query \credit remotely only in top-K benchmarks because it performs regression and cannot be cascaded.
We show all benchmarks' transformation graphs in
Figure \ref{fig:transformation_graphs}.

\subsection{Evaluating \willump}\label{sec:eval-e2e}

\paragraph{\willump Throughput.}

\begin{figure}[t]
	\includegraphics[width=\linewidth]{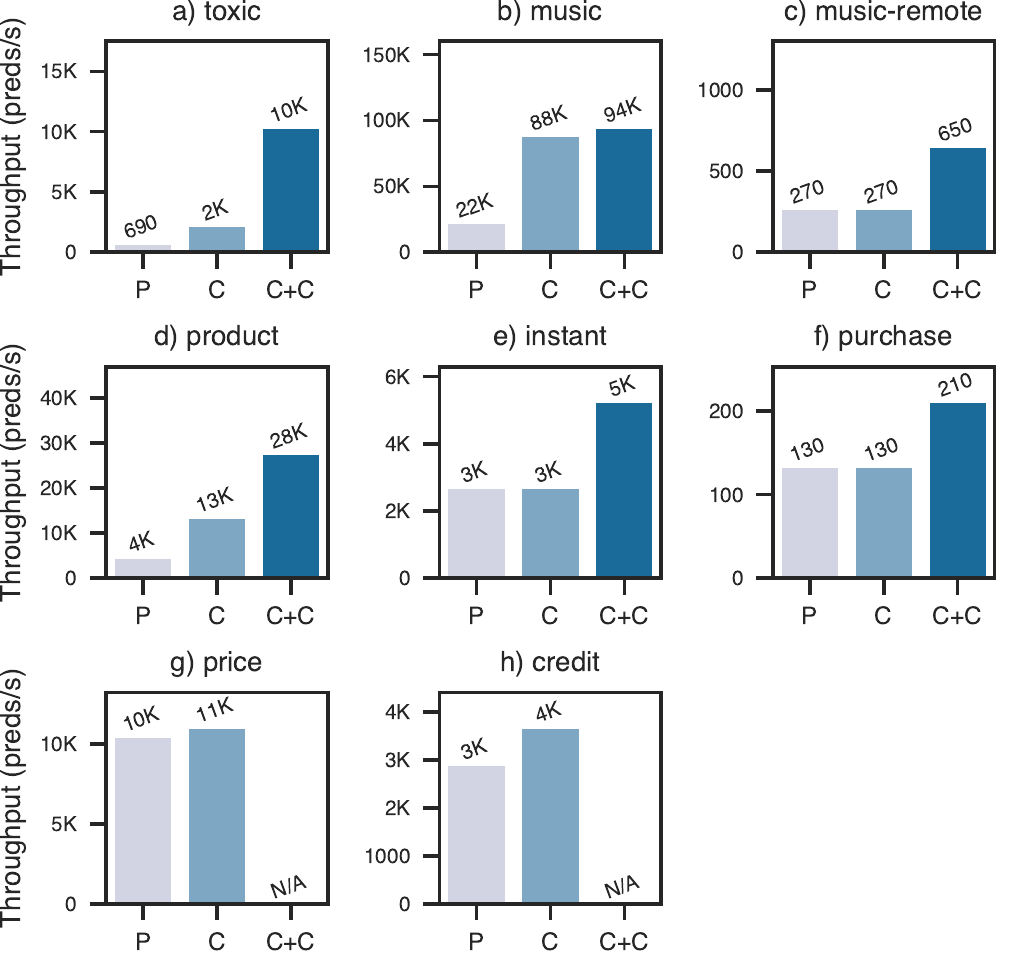}
	\caption{\willump performance on offline batch queries. P means (unoptimized) Python,
	C means compiler optimizations only, C+C means compiler and cascades optimizations.}
	\label{fig:batch_e2e}
	\squeezeup
\end{figure}

We first evaluate \willump on offline batch queries,
showing results in Figure \ref{fig:batch_e2e}.
First, we apply \willumps compiler optimizations.
These improve the performance of all compilable benchmarks
by up to 4.3$\times$.

Then, we apply end-to-end cascades to all classification benchmarks.  For all benchmarks, we set an accuracy target of 0.1\% less
than the accuracy of the original model,
but we did not observe a statistically significant change in accuracy for any benchmark.

End-to-end cascades improve benchmark performance by up to 5$\times$.
Interestingly, cascades are least effective on \music, which queries pre-computed features
from an in-memory database.  This is very fast when compiled, so feature computation accounts
for a small portion of overall runtime and potential gains from cascades are limited.  
In \musicremote, we moved the features to a remote database, so querying
them was more costly and cascades became more effective, providing a 2.4$\times$ speedup.

\paragraph{\willump Latency.}\label{sec:eval-point}

\begin{figure}[t]
	\includegraphics[width=\linewidth]{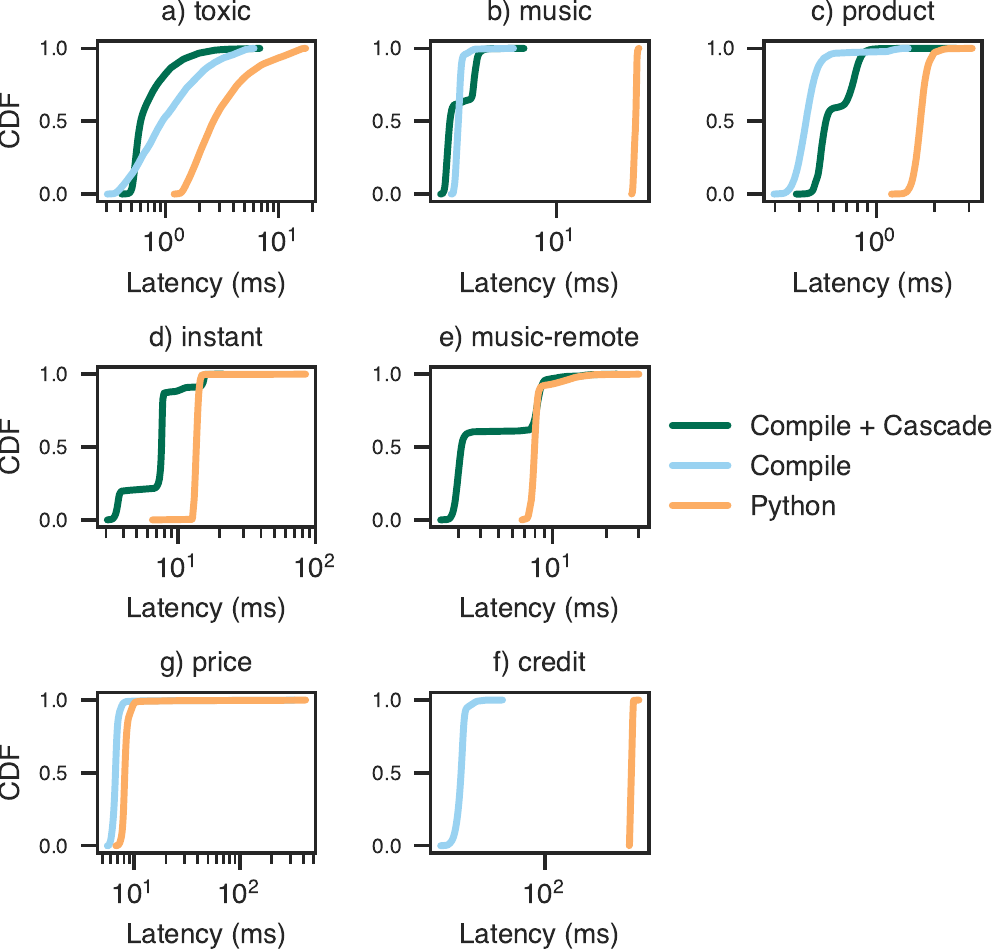}
	\caption{\willump latency CDFs on online point queries, with one outstanding query at a time.  Benchmarks in the second row contain no compilable operators;
	we only apply cascades.  Benchmarks in the third row do not perform classification;
	we only apply compiler optimizations.}
	\label{fig:latency_cdf}
	\squeezeup
\end{figure}

We next evaluate \willump on online point queries,
showing results in Figure \ref{fig:latency_cdf}.
We evaluate all benchmarks except \purchase; the Python implementation of the deep feature synthesis 
algorithm was not designed for low-latency queries.
When evaluating, we make one query at a time; each query contains one data input.
We first apply \willumps end-to-end compiler optimizations to all compilable benchmarks.
These decrease p50 and p99 latency by 1.3-400$\times$.
The large speedup on \music and \credit is
enabled by \willumps low-latency Weld drivers, which call into \willumps Weld code far faster than
the benchmarks' original Pandas implementations can call into their underlying C code;
this is critical for point operations.

We then apply end-to-end cascades to all classification benchmarks.
These improve p50 latency by up to 2.5$\times$ for most benchmarks.
However, because cascades only speed up prediction of some
data inputs (those classified by the approximate model), they
do not improve p99 latency.  Cascades are least effective on \product
and \music; in both cases this is because the latency contribution
of feature computation is small compared to that
of the model.  We demonstrate this with an alternate version of \music,\,
\musicremote, which queries features from remote databases instead of an in-memory store.
As in the previous section, more expensive feature computation makes cascades more effective;
they decrease p50 latency of \musicremote by 2.5$\times$.

\squeezeup
\paragraph{Top-K Queries.}\label{sec:eval-topK}

\begin{figure}[t]
	\includegraphics[width=\linewidth]{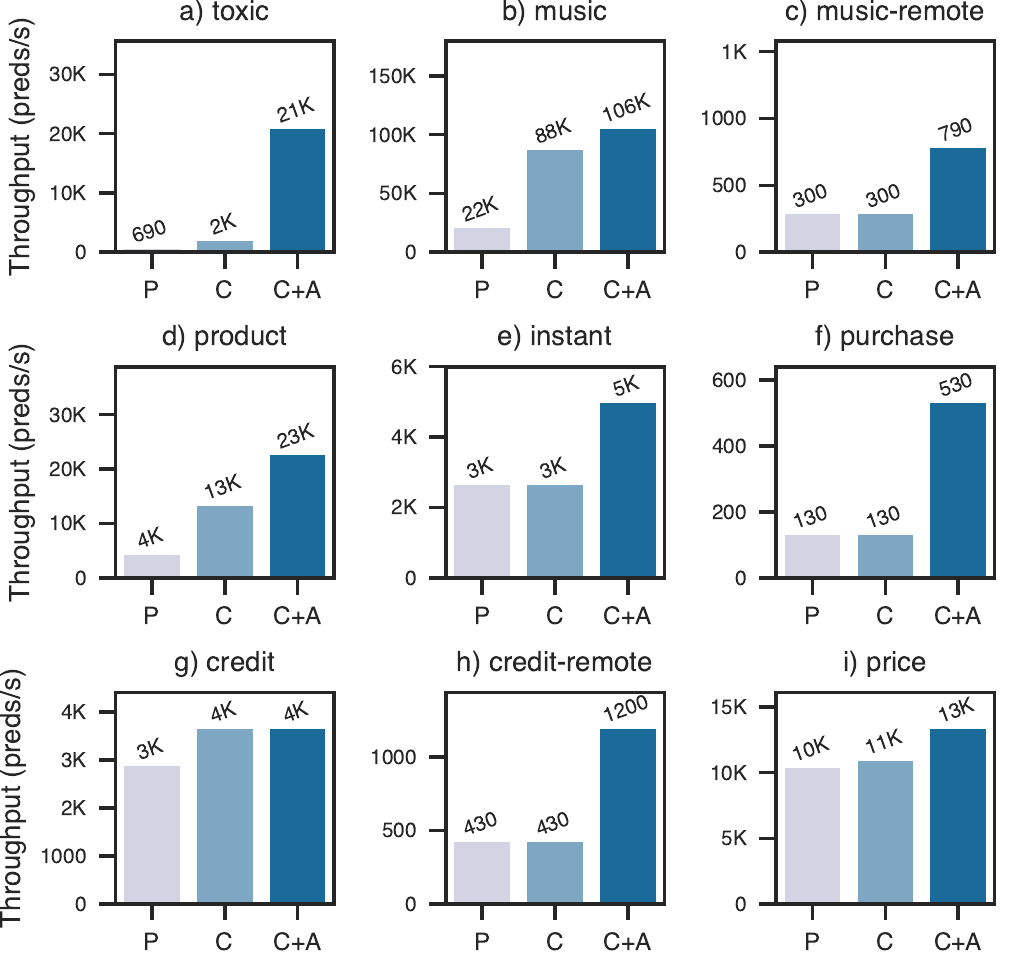}
	\caption{\willump performance on top-K queries. P means (unoptimized) Python,
		C means compiler optimizations only, C+A means compiler and top-K approximation optimizations.}
	\label{fig:topk_results}
	\squeezeup
\end{figure}

We now evaluate \willump on top-K queries, showing results in Figure \ref{fig:topk_results}.
We use $K=20$, query over the entire validation set, and set a minimum
precision of 0.95.

We first apply \willumps end-to-end compiler optimizations to compilable benchmarks;
these perform the same as in the batch setting.  We then apply \willumps top-K query approximation
optimization.  This produces performance improvements ranging from 1.3-10$\times$ with precision
always above the minimum.  Smaller speedups occur in benchmarks with relatively expensive models, such as
\music, \credit, and \price, as well as in benchmarks where differences between scores of high-scoring candidates were small
(less than a hundredth of a percent), like \product.
Benchmarks with less expensive models and more differentiation between high-scoring candidates, like \musicremote, \creditremote,
and \toxic, have larger speedups.

\squeezeup
\paragraph{Integration with Clipper.}\label{sec:eval-clipper-comparison}

\begin{figure}[t]
	\includegraphics[width=\linewidth]{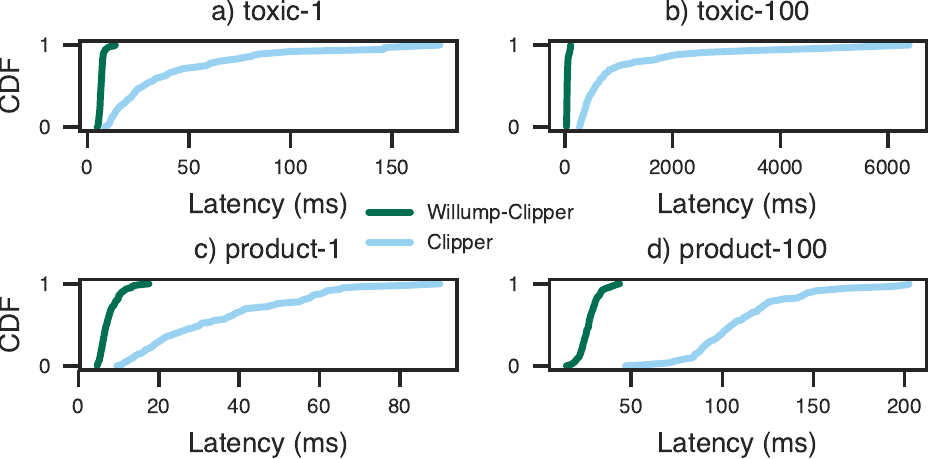}
	\caption{End-to-end latency CDFs on queries made using Clipper with and without \willump optimization at batch sizes of 1 and 100.}
	\label{fig:clipper-results}
	\squeezeup
\end{figure}

We next evaluate integration of \willump with Clipper,
showing results in Figure \ref{fig:clipper-results}.
We optimize \product and \toxic with \willumps compiler and cascades optimizations
and serve with Clipper, evaluating end-to-end query latencies.
At a batch size of 1, \willump improves p50 latency by 2.5-3$\times$ and p99 latency by 4.5-12$\times$.
At a batch size of 100, \willump improves p50 latency by 4-9$\times$ and p99 latency by 5-25$\times$.
These speedups are slightly smaller than in prior experiments but increase with batch size because Clipper has significant overheads,
including serialization, RPC processing time, etc.

\squeezeup
\paragraph{\willump Overhead.}

Finally, we evaluate the training times of \willumps end-to-end cascades and top-K query approximation optimizations,
showing results in Table~\ref{table:training-times}.
\willumps end-to-end training times are 1.0-6.9$\times$ as long as the end-to-end
training times of the original pipelines.
The increase in training time is largest for benchmarks whose end-to-end training time is dominated
by model training and comes largely from the need to train several approximate models during approximate model feature selection.
We believe \willumps tradeoff of train performance for inference performance is acceptable
because for many real ML applications, training consumes far fewer resources than inference~\cite{hazelwood2018applied}.

\begin{table}[t]
	\resizebox{\columnwidth}{!}{%
		\begin{tabular}{lrrrrrrr}
			\toprule
			& Toxic & Music & Product & Instant & Purchase & Price & Credit \\ \midrule
			Cascades & 1.1$\times$ & 6.8$\times$ & 2.2$\times$ & 1.0$\times$ & 1.7$\times$ & N/A & N/A \\
			Top-K & 1.1$\times$ & 6.9$\times$ & 1.8$\times$ & 1.0$\times$ & 1.6$\times$ & 3.8$\times$ & 3.9$\times$ \\ 
			\hline
		\end{tabular}%
	}
	\caption{Ratios of \willump to Python training times when training end-to-end cascades and top-K query approximations.}
	\label{table:training-times}
	\squeezeup
\end{table}
	\section{Related Work}

\paragraph{Model Serving:}
Researchers and commercial vendors have developed many model serving systems.
Some are general-purpose, serving different models from different frameworks.
These include research systems, such as Clipper \cite{crankshaw2017clipper} and Rafiki \cite{wang2018rafiki},
and commercial platforms, such as Amazon's Sagemaker and Microsoft's Azure ML,
among others~\cite{tfseving_2019, predictionIO}.
These systems aim to reduce the difficulty of deploying ML models.
Typically, they consider models to be black boxes and implement only
pipeline-agnostic optimizations such as end-to-end caching and
adaptive batching \cite{crankshaw2017clipper}.
However, recently researchers have developed more powerful optimizations,
such as the use of erasure codes to improve accuracy in the face of unavailability~\cite{kosaian2019parity}.
As an optimizer for ML inference pipelines,
\willump synergizes with general-purpose model serving systems,
significantly improving their performance with its statistically-aware optimizations, as
we show in Section \ref{sec:eval-clipper-comparison}.

Other model serving systems are application-specific.
For example, Noscope \cite{kang2017noscope} and Focus \cite{hsieh2018focus} improve performance
of neural nets querying large video datasets.
Major web companies have also developed specialized systems for video recommendation \cite{davidson2010youtube}
and ad-targeting \cite{cheng2016wide}, among other tasks.
Also related is LASER~\cite{agarwal2014laser}, designed for linear models used in advertising, which provides latency guarantees by dropping slow-to-compute features.
\willump generalizes ideas from these application-specific systems, such as their use of cascades.

\squeezeup
\paragraph{Cascades:}

Cascades were initially developed for rapid object detection by applying
more complex classifiers to more object-like regions of an image~\cite{viola2001rapid}.
They have been widely applied to image and video
tasks such as pedestrian detection \cite{cai2015learning} and face recognition \cite{sun2013deep}.
Some application-specific model serving systems, such as NoScope and Focus,
utilize cascades.

\squeezeup
\paragraph{Cost-Sensitive Training Algorithms:}
ML researchers have proposed many algorithms for incorporating feature cost into model training, similar
to \willumps cascades.  Prominent examples include \citet{xu2014classifier}, \citet{wang2011cascade}, and 
\citet{raykar2010designing}.  However, unlike \willump, these systems are neither general nor automatic.  
They perform no graph or dataflow analysis, assume all features are computationally independent,
require users to provide costs for all features, and restrict users to specific types of models
(such as linear models for \citet{wang2011cascade} or cascades of CART trees for \citet{xu2014classifier}).

\squeezeup
\paragraph{Top-K Approximation:}

Database researchers have proposed many algorithms for optimizing and approximating top-K queries.
These algorithms, including  
MPro~\cite{chang2002minimal} and approximate TA~\cite{theobald2004top}, among others~\cite{ilyas2008survey},
score candidates by querying one data source at a time; they drop candidates whose probability
of scoring in the top K given the queried data sources is too low.
However, they do not translate well to ML top-K queries.
They assume that scoring functions execute on a set of data sources which can be queried independently,
but ML models execute on features which are often computationally interdependent.
They also assume that scoring is done by a monotonic and usually linear aggregation function,
but ML models are typically nonlinear and offer no monotonicity guarantees.

Some systems for performing ML top-K queries use a \textit{retrieval model} to select high-scoring
inputs to rank with a more powerful model~\cite{cheng2016wide}.  This is similar to \willumps use of an
approximate model, but retrieval models are typically manually constructed while \willump constructs approximate
models automatically.
Therefore, to the extent of our knowledge, \willump is the first system to automatically optimize
ML top-K queries.

\squeezeup
\paragraph{ML Optimizers and Compilers:}

Several prior systems have developed optimizers and compilers for ML workloads.
Unlike \willump, many specialize in improving neural network performance.
For example, TVM~\cite{chen2018tvm} compiles deep neural nets to different architectures
and NVIDIA TensorRT~\cite{moroney_sharma_2018} is a library optimized for fast inference performance on NVIDIA GPUs.
Additional optimization techniques include knowledge distillation \cite{hinton2015distilling}
and approximate caching \cite{kumar_2019}.
Some systems, like \willump, optimize ML pipeline performance.
For example, KeystoneML~\cite{sparks2017keystoneml} optimizes distributed training pipelines.
Closer to \willump is Pretzel~\cite{lee2018pretzel}, which improves ML inference
performance through end-to-end compiler optimizations such as loop fusion and vectorization.
These are similar to the optimizations \willump implements through Weld compilation,
but unlike \willump Pretzel does not implement statistical optimizations such as cascades or top-K approximation.

	\section{Conclusion} \label{sec:conclusion}

This paper presents \willump, a statistically-aware end-to-end optimizer for ML inference.
\willump leverages unique properties of ML inference applications to automatically
improve their performance through statistical optimizations,
such as end-to-end cascades and top-K query approximation.
\willump improves the performance of real-world ML inference pipelines by up to an order of magnitude
over existing systems.

\section{Acknowledgments}

This research was supported in part by affiliate members and other supporters of the Stanford DAWN project---Ant Financial, Facebook, Google, Infosys, NEC, and VMware---as well as Toyota Research Institute, Northrop Grumman, Cisco, SAP, and the NSF under CAREER grant CNS-1651570. Any opinions, findings, and conclusions or recommendations expressed in this material are those of the authors and do not necessarily reflect the views of the National Science Foundation.
Toyota Research Institute (``TRI'') provided funds to assist the authors with their research but this article solely reflects the opinions and conclusions of its authors and not TRI or any other Toyota entity.
	
	\bibliography{willump-base}
	\bibliographystyle{sysml2019}
	
	\appendix

\section{Microbenchmarks}
\label{sec:microbenchmarks}

In this section, we analyze the behavior and overhead of \willumps optimizations in more detail.
\squeezeup
\paragraph{Cascades Tradeoffs.}

\begin{figure}[t]
	\includegraphics[width=\linewidth]{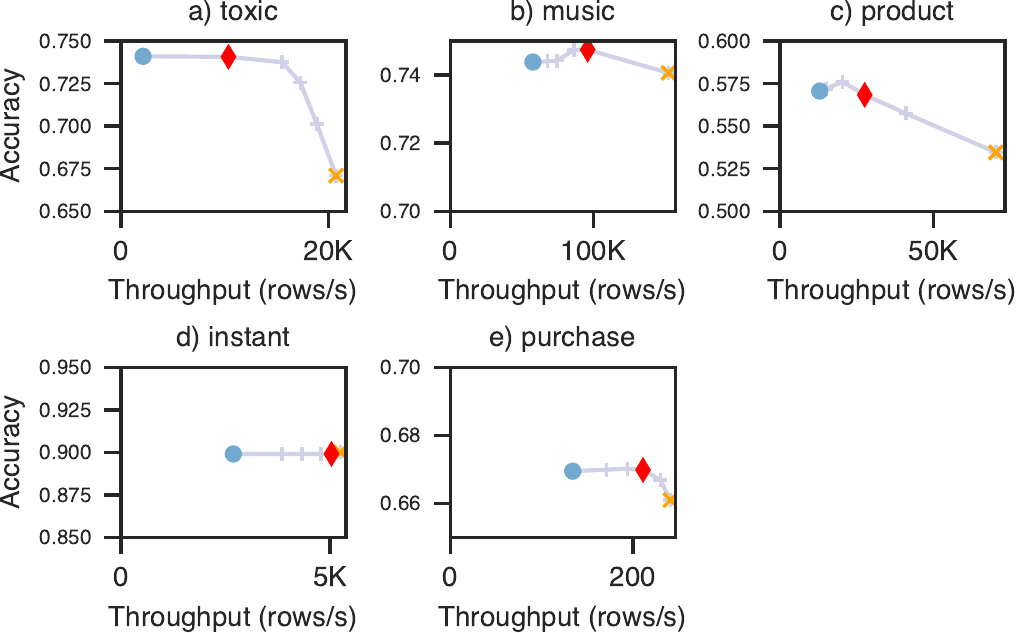}
	\caption{Throughput versus accuracy on all classification benchmarks with varying cascade thresholds.
		A blue circle marks performance of the original model, an orange X marks performance of the approximate model,
		and a red diamond marks \willumps chosen cascade threshold.}
	\label{fig:cascades-tradeoffs}
\end{figure}

We first examine the behavior of \willumps cascades optimization.
For each classification benchmark, we graph performance and accuracy at
varying cascade thresholds in Figure \ref{fig:cascades-tradeoffs}.
On each graph, a blue circle marks the performance of the original model and an orange X that of the approximate model;
points in between are cascaded models with varying cascade thresholds.
As a reminder, the cascade threshold is the confidence the approximate model must have in a prediction to
return the prediction and not cascade to the original model.

For all benchmarks, cascaded models with high cascade thresholds are faster than
the original model but have similar accuracy.  \willumps cascades algorithm automatically chooses these cascade thresholds
to maximize performance without statistically significant accuracy loss.  On each graph, the point marked
with a red diamond is the chosen threshold.

For most benchmarks, as the cascade threshold decreases,
performance continues to improve but accuracy falls off.
This shows cascades are working as intended.
At high cascade thresholds, the approximate model classifies easy data inputs and the
original model classifies hard data inputs, so accuracy is high.
At low thresholds, the approximate model must classify hard data inputs, so accuracy falls.

\paragraph{Compilation Times}

We also evaluate \willumps pipeline compilation times.  We find these rarely exceed thirty seconds.
The exceptions are benchmarks which query in-memory data stores (\music and \credit);
it takes \willump several minutes to optimize them as it must convert the data stores
into a format which Weld can query.
%
%
%
%
%

\section{Artifact Appendix}

\subsection{Abstract}

This artifact contains the implementation of the \willump optimizer described in this paper.  
Furthermore, this artifact contains the scripts and datasets needed to reproduce the key performance
results in this paper, specifically the batch, point, and top-K performance comparisons in
Figures \ref{fig:batch_e2e}, \ref{fig:latency_cdf}, and
\ref{fig:topk_results}.

\subsection{Artifact check-list (meta-information)}

{\small
\begin{itemize}
  \item {\bf Program: } The \willump optimizer implemented in Python.
  \item {\bf Data set:} Competition datasets for each benchmark, all provided or linked to.
  \item {\bf Run Time Environment:}  Python 3 (Tested with Python 3.6.8).
  \item {\bf Metrics: } For batch (Figure \ref{fig:batch_e2e}) and top-K (Figure \ref{fig:topk_results}) experiments, pipeline throughput.
  	For point (Figure \ref{fig:latency_cdf}) experiments, pipeline latency.
  \item {\bf Output: } Throughputs and latencies (p50 and p99) printed to stdout.
  \item {\bf Experiments: }  Run each benchmark in the batch, point, and top-K setting with no optimizations, compiler optimizations
  only, and compiler optimizations plus cascades or top-K approximation optimizations.
  \item {\bf Publicly available?: } Yes.
\end{itemize}

\subsection{Description}

\subsubsection{How delivered}

The artifact is hosted on GitHub at \url{https://github.com/stanford-futuredata/Willump}.
The artifact, its instructions, its benchmark scripts, and its datasets are publicly available.
Additionally, an archival version of the artifact is available at \url{https://doi.org/10.5281/zenodo.3687193}.

\subsubsection{Software dependencies}

Experiments were run using Ubuntu 18.04 with Python 3.6.8.
Full installation instructions including dependencies are
here: \url{https://github.com/stanford-futuredata/Willump/blob/master/README.md}.

\subsubsection{Data sets}

The datasets of the competitions our benchmarks were curated from were used.  All are included in the artifact
or linked to from \url{https://github.com/stanford-futuredata/Willump/blob/master/BENCHMARKS.md}.

\subsection{Installation}

Full installation instructions are here: \url{https://github.com/stanford-futuredata/Willump/blob/master/README.md}.

\subsection{Experiment workflow}

Full benchmark instructions are here:  \url{https://github.com/stanford-futuredata/Willump/blob/master/BENCHMARKS.md}.

\subsection{Evaluation and expected result}

Artifact evaluation is expected to reproduce the performance comparisons in Figures \ref{fig:batch_e2e}, \ref{fig:latency_cdf}, and
\ref{fig:topk_results} of the paper.  We expect the results of evaluation to show a similar performance trend as these figures.


\end{document}